\documentclass{appolb}
\usepackage{graphicx}
\usepackage{subfig}
\usepackage{floatrow}
\usepackage{hyperref}       
\usepackage{url} 
\usepackage{amsmath}

\def\Acknowledgements{\bigskip  \bigskip \begin{center} \begin{large}
             \bf ACKNOWLEDGEMENTS \end{large}\end{center}}

\begin{document}
\title{Comparison of Machine Learning Approach to other Commonly Used Unfolding Methods%
\thanks{Presented at XXVII Cracow Epiphany Conference, LHC Physics: Standard Model and Beyond}%
}
\author{Petr Baro\v n
\address{Joint Laboratory of
Optics of Palacký University and Institute of Physics AS CR, Faculty of
Science,
Palacky University, 17. listopadu 12, 771 46 Olomouc, Czech~Republic\\ petr.baron@upol.cz }
\\
}
\maketitle
\begin{abstract}
Unfolding in high energy physics represents the correction of measured spectra in data for the finite detector efficiency, acceptance, and resolution from the detector to particle level.
Recent machine learning approaches provide unfolding on an event-by-event basis allowing to simultaneously unfold a large number of variables 
and thus to cover a wider region of the features that affect detector response. 
This study focuses on a simple comparison of commonly used methods in RooUnfold
package to the machine learning package Omnifold.

\end{abstract}
  
\section{Introduction}
\begin{figure}[h!]
\subfloat[]{\includegraphics[width=0.33\linewidth]{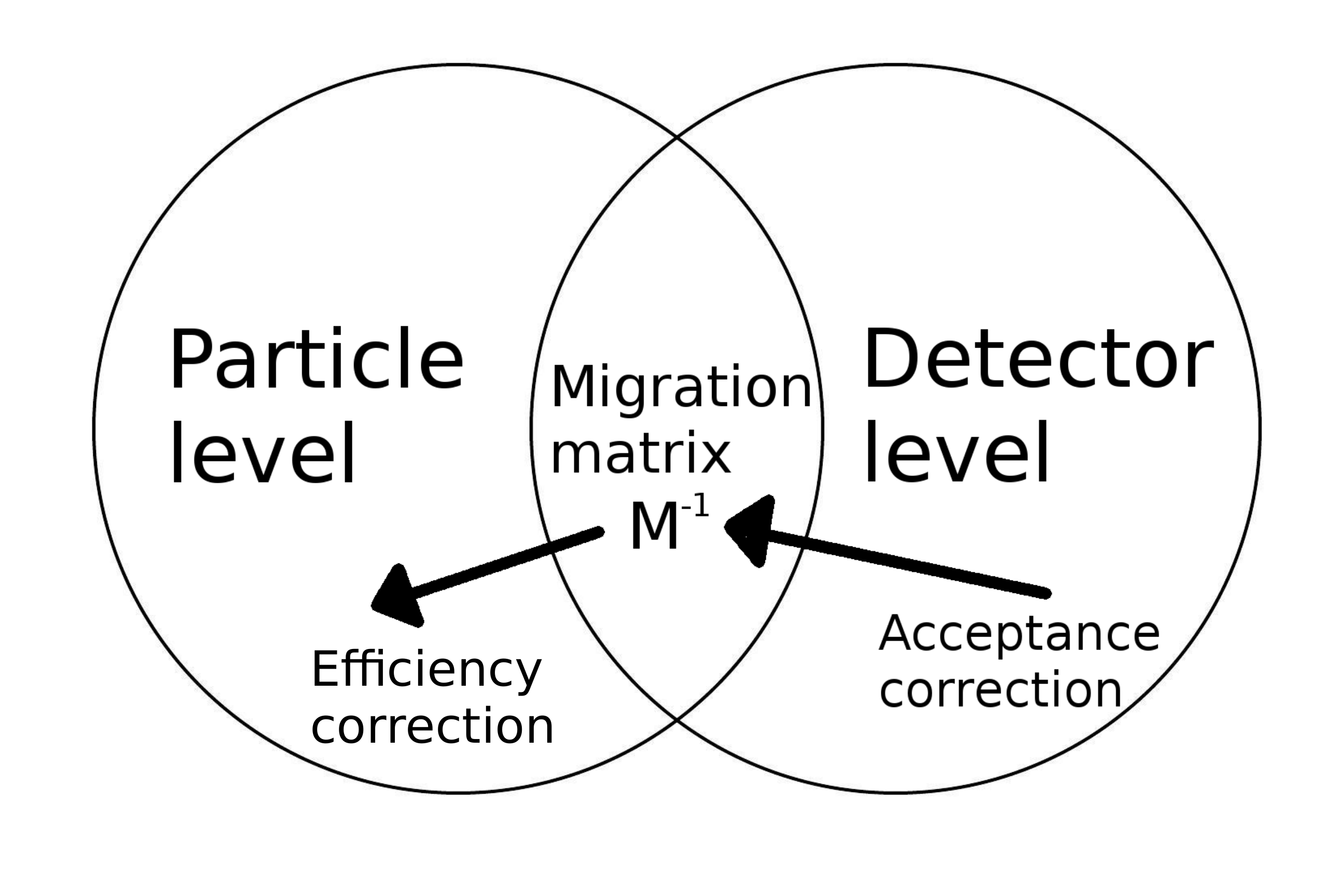}}
\subfloat[]{\includegraphics[width=0.33\linewidth]{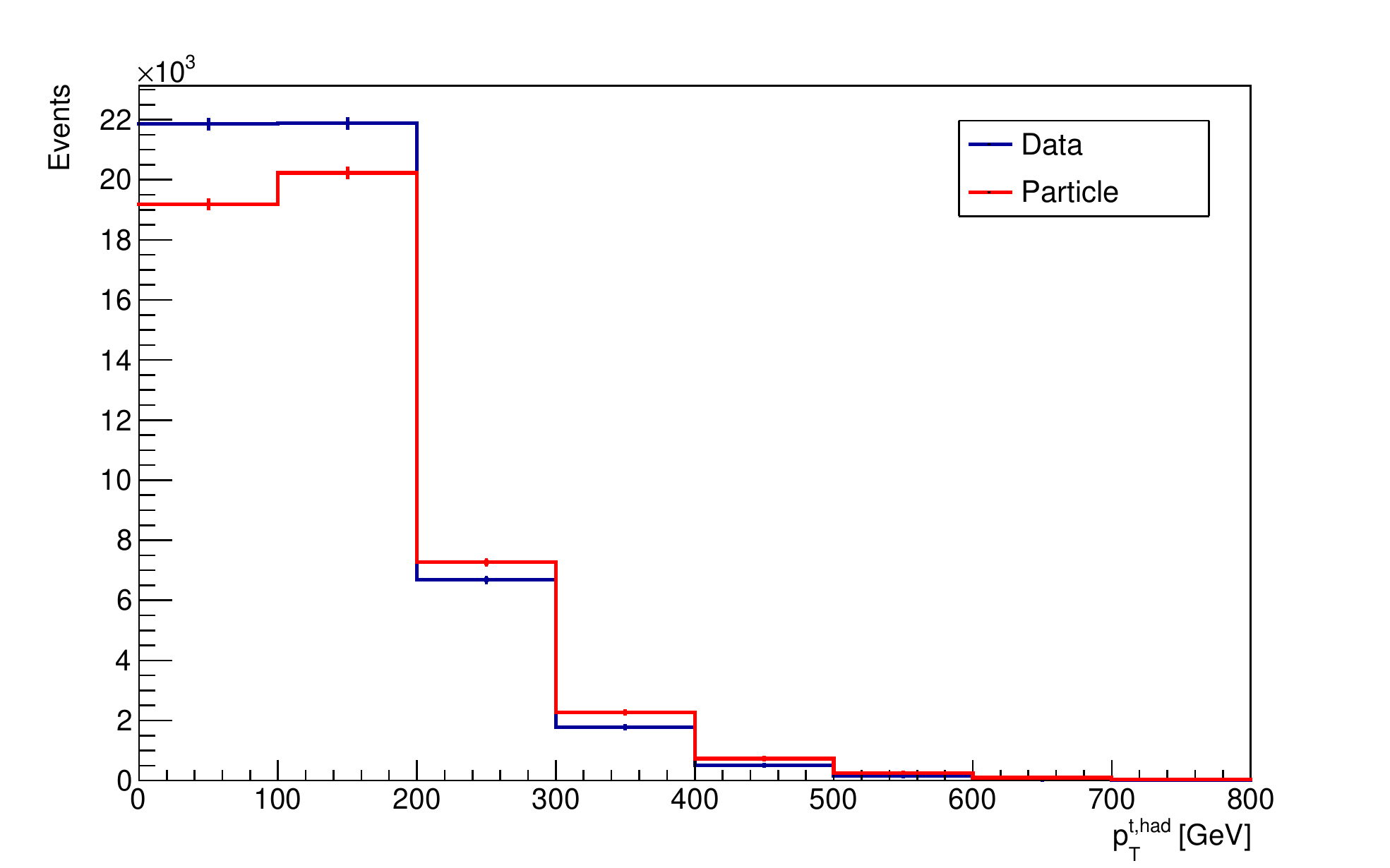}}\\
\subfloat[]{\includegraphics[width=0.33\linewidth]{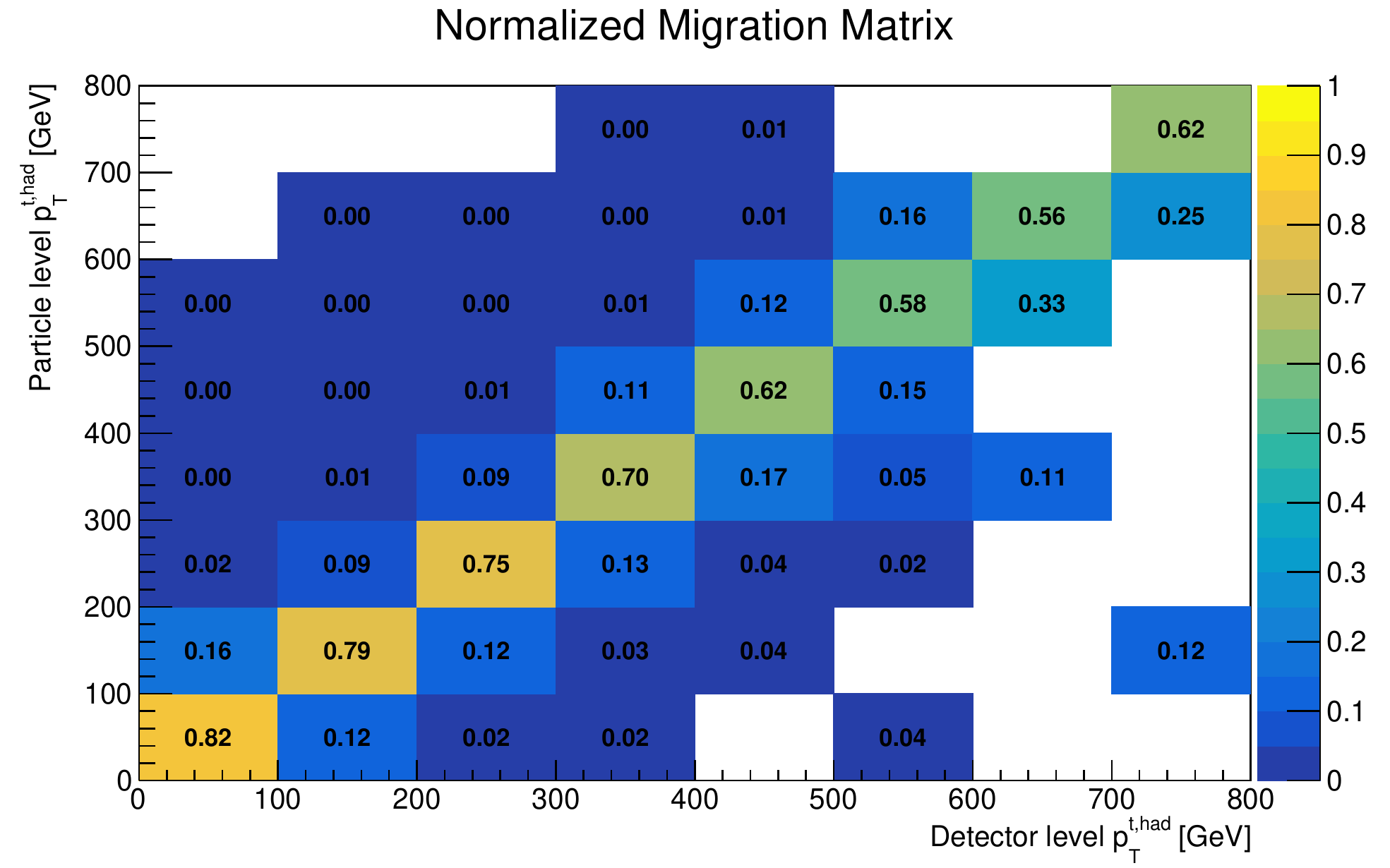}}
\subfloat[]{\includegraphics[width=0.33\linewidth]{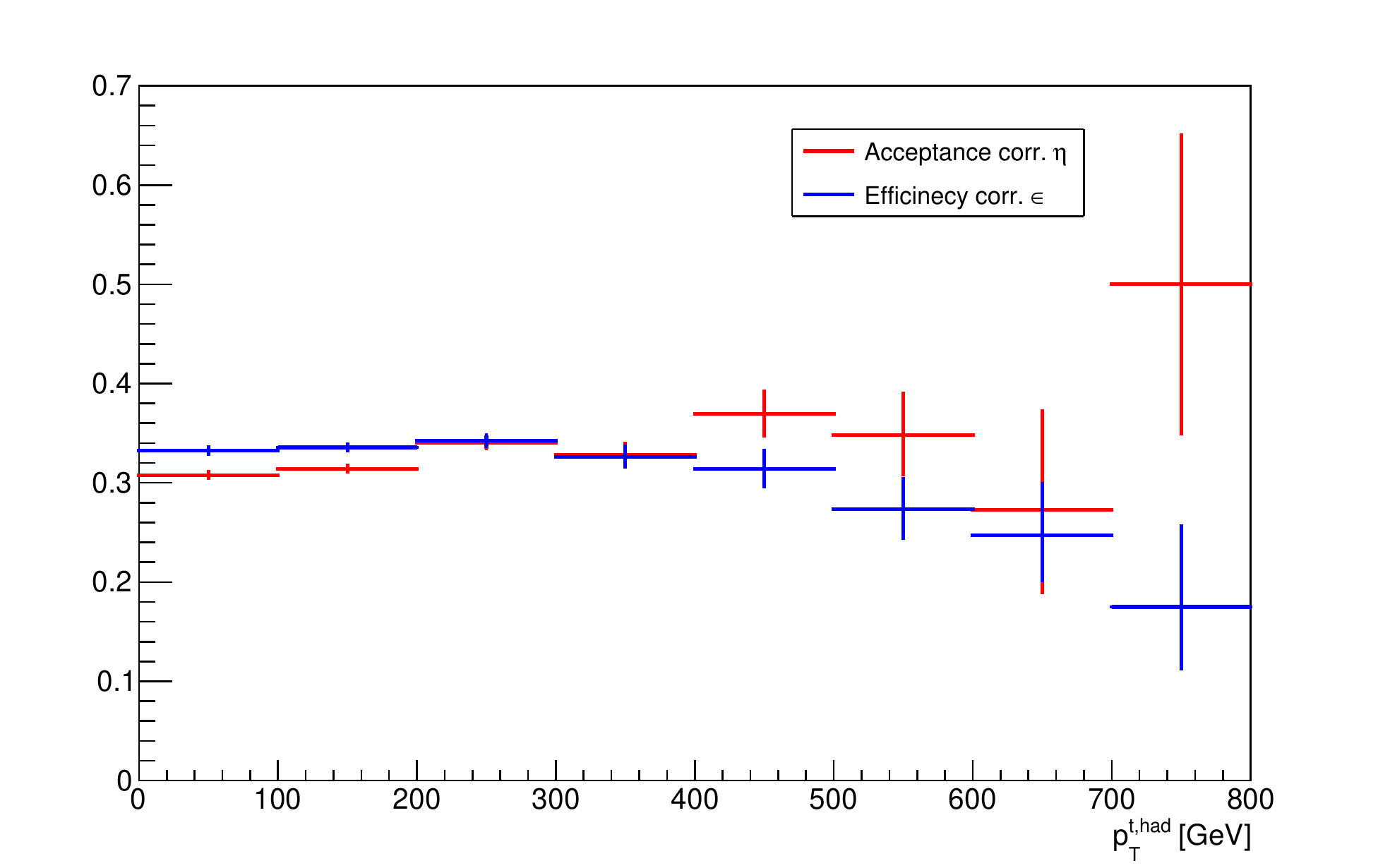}}
\caption{Unfolding inputs. \textbf{a)} Unfolding procedure diagram \textbf{b)} Detector-level (blue) and particle-level (red) spectra. \textbf{c)} Migration matrix between particle and detector levels. \textbf{d)} Efficiency (blue) and acceptance (red) corrections as~a function of the transverse momentum of the hadronically decaying top quark.~\cite{Baron:2020auz}}
\label{fig:ingr}
\end{figure}

The equation of unfolding can be written as 
\begin{equation}
    p = \frac{1}{\epsilon}\cdot M^{-1}\cdot\eta\cdot(D - B);~\cite{Symmetry}
\end{equation}
where $D$ is the data spectrum from which the background spectrum $B$ is 
subtracted followed by multiplication of acceptance correction $\eta$ so the 
main input to the unfolded procedure is prepared. The unfolding is here 
schematically given by a so-called migration matrix $M^{-1}$ which maps 
one-to-one events from the detector to particle level. Behind the symbol, $M^{-1}$ 
one could also imagine not necessarily the algebraic matrix inversion, but rather different 
unfolding methods, because the treatment of unfolding input differs. However,
the aim of this study is not to fully describe all the methods separately, but 
rather to make a comparison between them. The result of unfolding has to be 
corrected by the detector efficiency $\epsilon$ to obtain the unfolded 
"truth" spectrum $p$ ideally close to the truth (particle) level. The background $B$ is not considered in this study.
Figure~\ref{fig:ingr} provides insight to 
the ingredients on the example of the transverse momentum spectrum of the 
hadronically decaying top quark in process $pp\rightarrow t\bar{t}$~\cite{Symmetry}.

\section{Machine Learning Approach}
The machine learning unfolding is similar to e.g. face recognition problem. 
The idea is to take all the possible information from the detector in a similar way as a  
photo and train some neural network to classify what process 
returns such a signature in the detector as the photo of the face classifies one particular person.\par
\begin{figure}[h!]
  \includegraphics[width=0.6\linewidth]{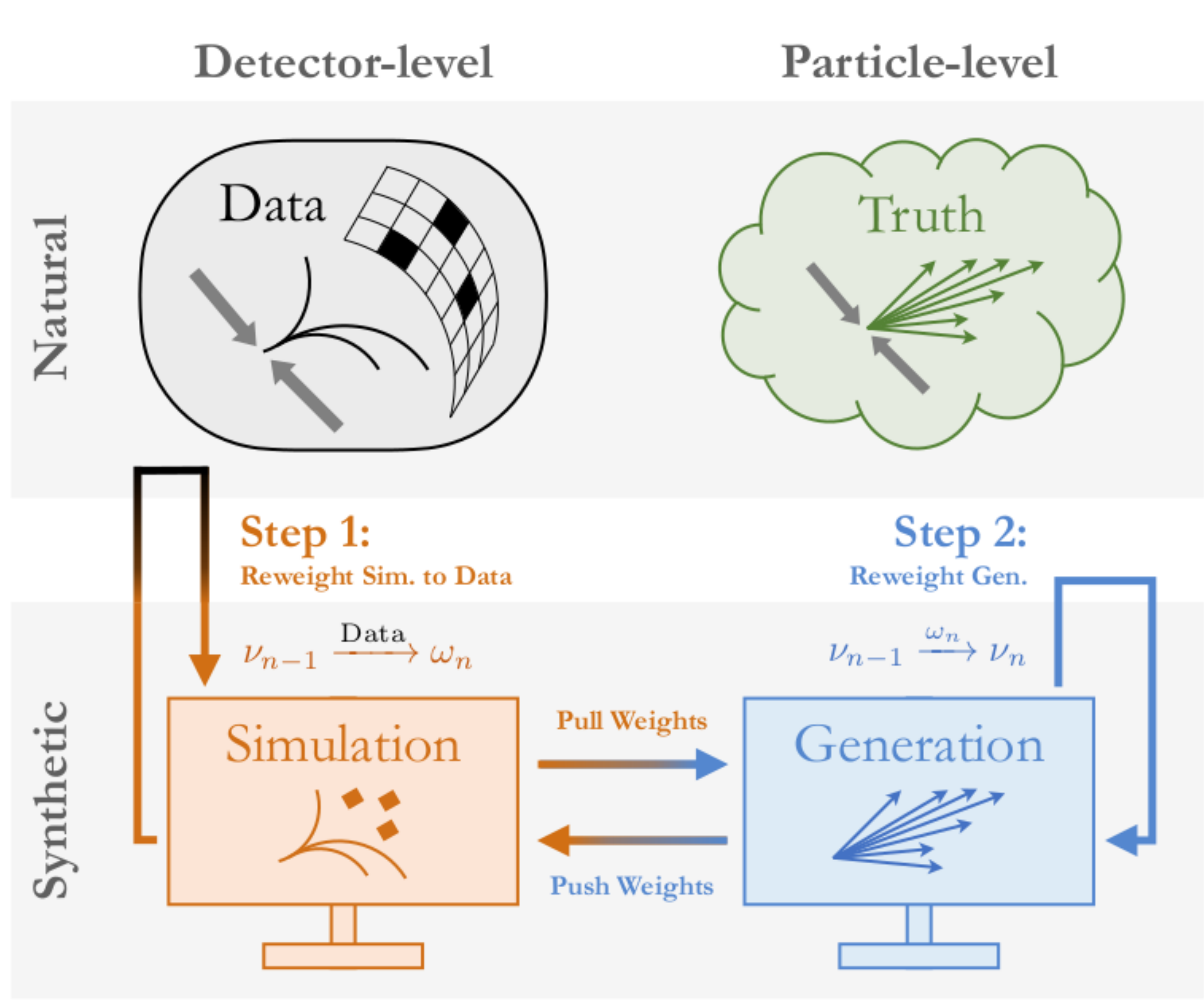}
  \caption{An illustration of the OmniFold method applied to a set of synthetic 
  and natural data. As a first step, starting from prior 
  weights $\nu_0$, the detector-level synthetic data (“simulation”) is 
  reweighted to match the detector-level natural data (simply “data”). 
  These weights $\omega_1$ are pulled back to induce weights
  on the particle-level synthetic data (“generation”). As a second step, 
  the initial generation is reweighted to match the new 
  weighted generation. The resulting weights $\nu_1$ are pushed forward 
  to induce a new simulation, and the process is iterated~\cite{Omnifold}.}
  \label{fig:omni}
\end{figure}

In both cases, machine learning needs the truth information to train the neural 
network, e.g. the face on the photo belonging to a particular person or the detector signature 
belonging to the process $pp\rightarrow t\bar{t}$. The model describing the 
physics is still necessary. Authors of Omnifold~\cite{Omnifold} call the truth process and truth detector signature 
natural and the model is called synthetic, see Figure~\ref{fig:omni}.
\par
The authors extended the idea of Iterative Bayes unfolding to continuous form and 
with machine learning concept enabled to perform unfolding event-by-event so the trained
network returns a set of weights for each event or a function which can be applied to measured data. 
The detailed description of the algorithm is in Appendix of~\cite{Omnifold}.
\section{Performing Closure Test}\label{sec:closure}
The closure test of the unfolding method is to unfold not the measured data, but 
rather the generated particle level spectrum. If the method is consistent their ratio should be 
close to unity. To avoid other systematical uncertainties from efficiency and 
acceptance corrections the events are chosen only from the overlap of the particle and detector 
level phase spaces, see the intersection of the circles in Figure~\ref{fig:ingr}.\par
\begin{figure}[h!]
  \subfloat[]{\includegraphics[width=0.33\linewidth]{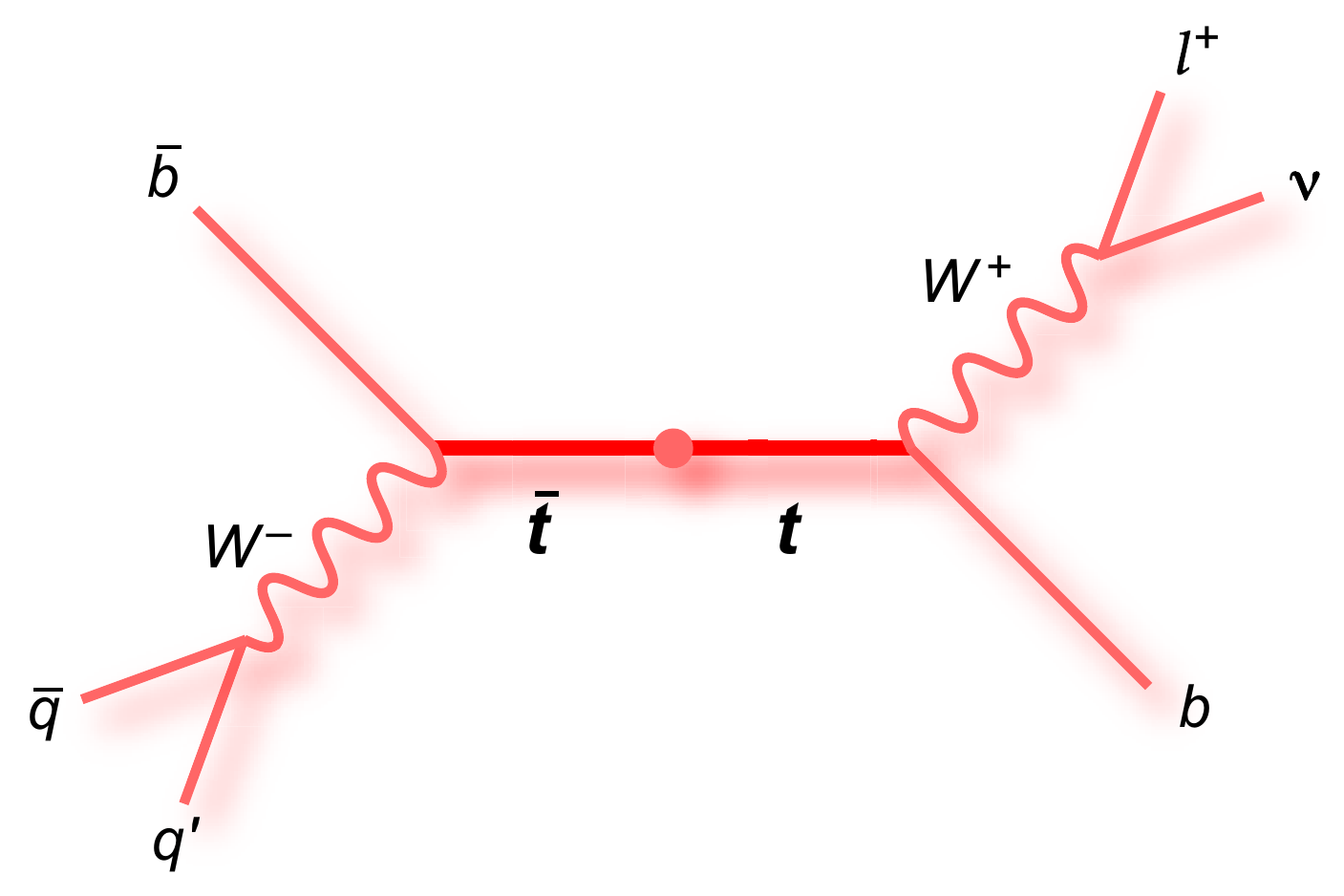}}
  \subfloat[]{\includegraphics[width=0.33\linewidth]{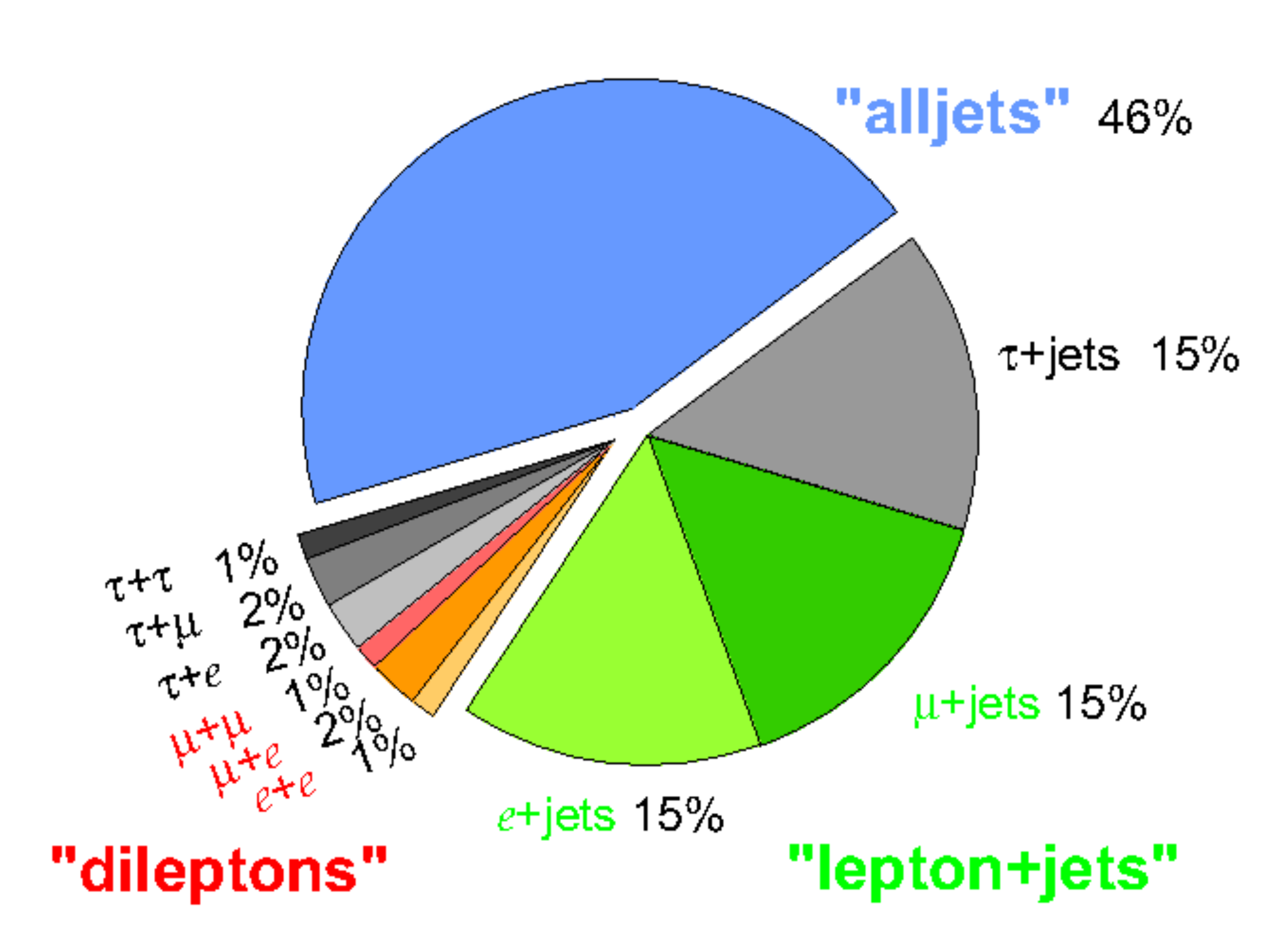}}\\
  \caption{\textbf{a)} Final state diagram of the process 
  $pp\rightarrow t\bar{t}$ \textbf{b)} Top quark pair production branching ratios.}
  \label{fig:ljets}
\end{figure}

As the process of study, the process of top quark pair production in proton-proton 
collisions $pp\rightarrow t\bar{t}$ in $\ell +$ jets channel was chosen, see Figure~\ref{fig:ljets}, simulated using MadGraph~\cite{madgraph} software with 
a generation of events using Pythia8~\cite{pythia} with the detector-level and the pseudo data 
simulated using Delphes~\cite{delphes} with ATLAS detector card. The basic selection and cuts were 
applied to obtain spectra with a similar shape to those measured at Large Hadron Collider (LHC) in the real ATLAS 
experiment, although comparison of unfolding methods could be performed with an arbitrary  
measured process. 

\par

The input data set was divided into two statistically independent sets, so the classical 
unfolding methods were using migration matrix build from one set and the input to unfolding 
procedure from the other set. The same exclusive sets were used for training the neural network and to perform 
the machine-learning unfolding. 

\section{Results}
The following spectra of interest were chosen: the transverse-momentum, mass, energy, and pseudo-rapidity of 
the hadronically, the leptonically decaying top quark, and also the $t\bar{t}$ system. The distribution $\phi$ 
was omitted due to its flat shape. 
In total $4\times3 = 12$ binned spectra were unfolded classically 
using the RooUnfold package~\cite{RooUnfold} with Bayes (3 iterations)~\cite{bayes}, SVD (k = 5)\cite{svd}, and Ids (k = 1)~\cite{ids} methods. 
The variables were used event-by-event in 
the neural network with 100 epochs to later perform simultaneous unfolding.\par
Even though results from the machine learning approach could be shown as continuous spectra, for 
comparison purposes particular fine binning was chosen. 

Particle-level spectra as the input into the unfolding procedure were chosen 
to perform the closure test. Thus the ratio between input particle-level spectrum and unfolded spectrum should 
be ideally close to unity as it was already discussed in Section~\ref{sec:closure}.\par
The metric of comparison is $\chi^2$ divided by the number of degrees of freedom NDF which 
is equal to the number of bins in the spectrum~\cite{Symmetry}.
\vspace*{-2cm}
\begin{figure}[h!]
  \subfloat[$p_T$]{\includegraphics[width=0.33\linewidth]{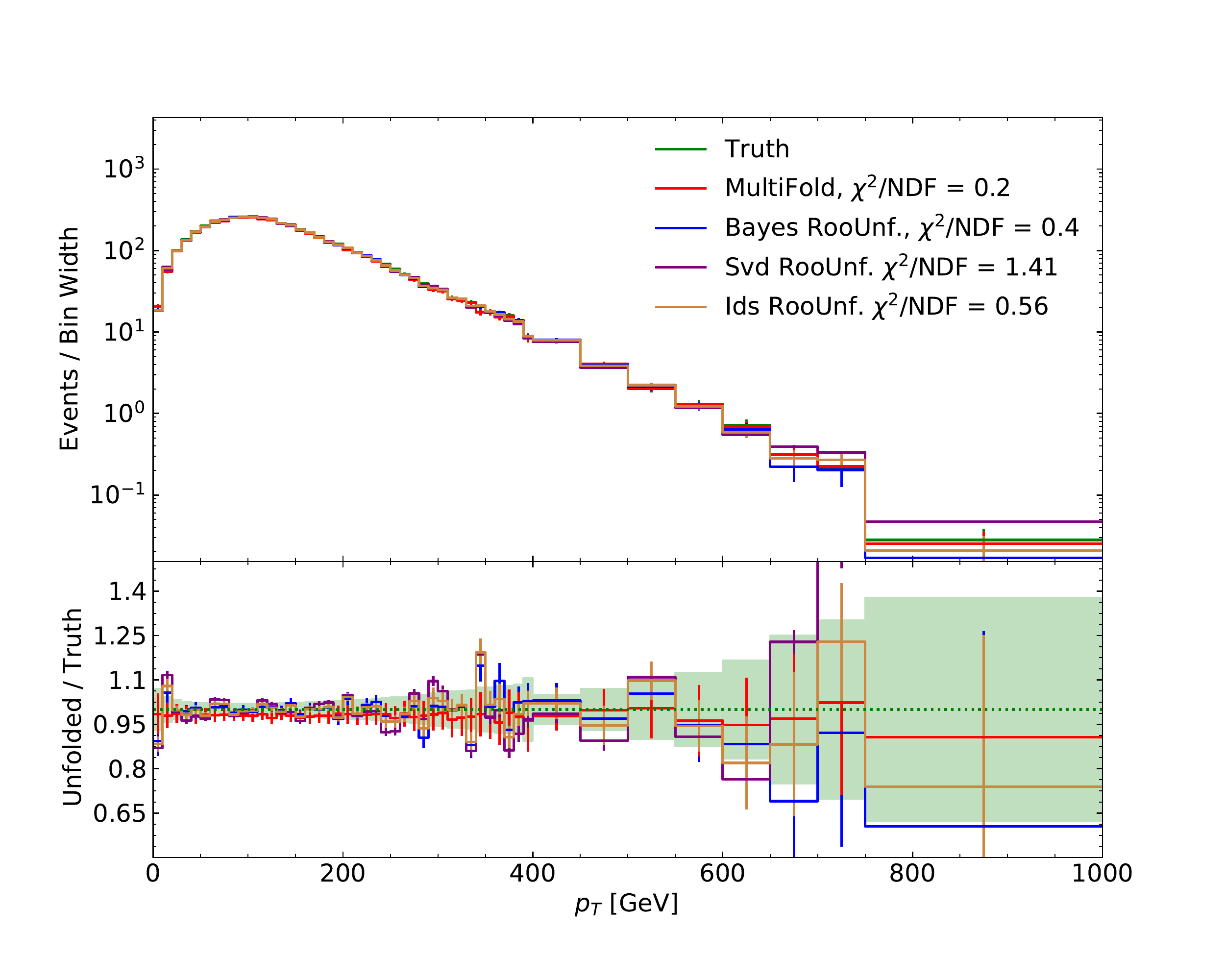}}
  \subfloat[$p_T$]{\includegraphics[width=0.33\linewidth]{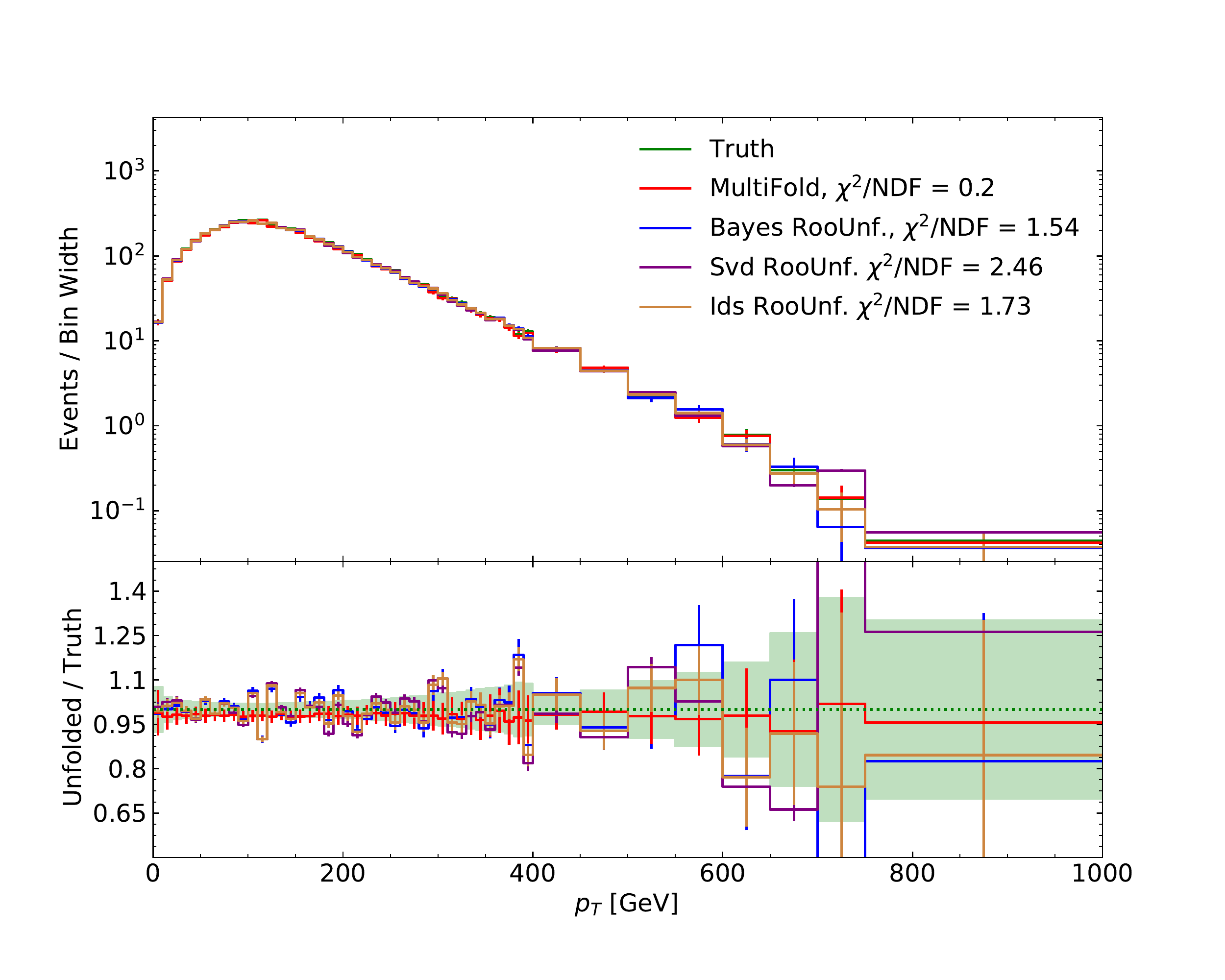}}
  \subfloat[$p_T$]{\includegraphics[width=0.33\linewidth]{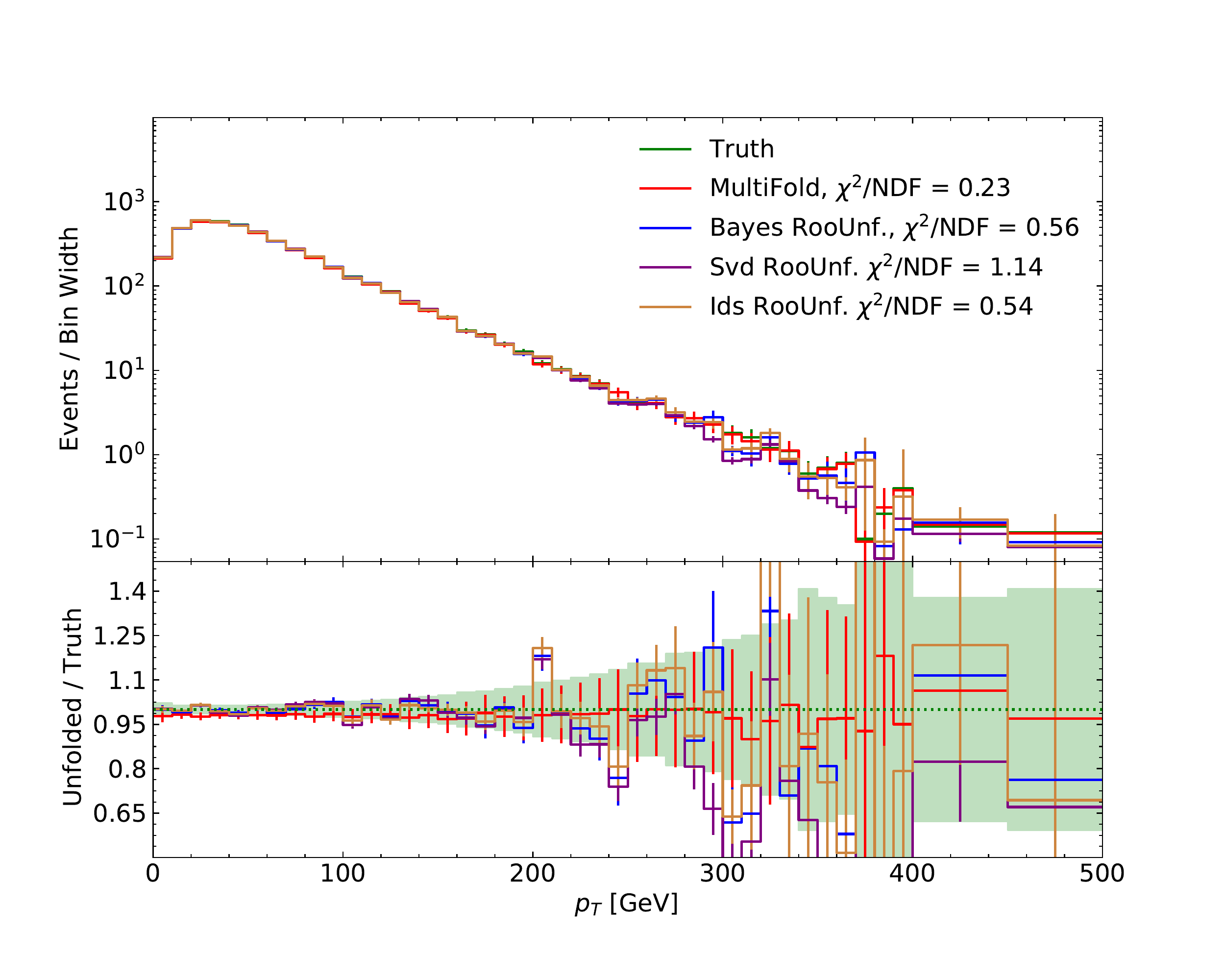}}\\

  \subfloat[Mass]{\includegraphics[width=0.33\linewidth]{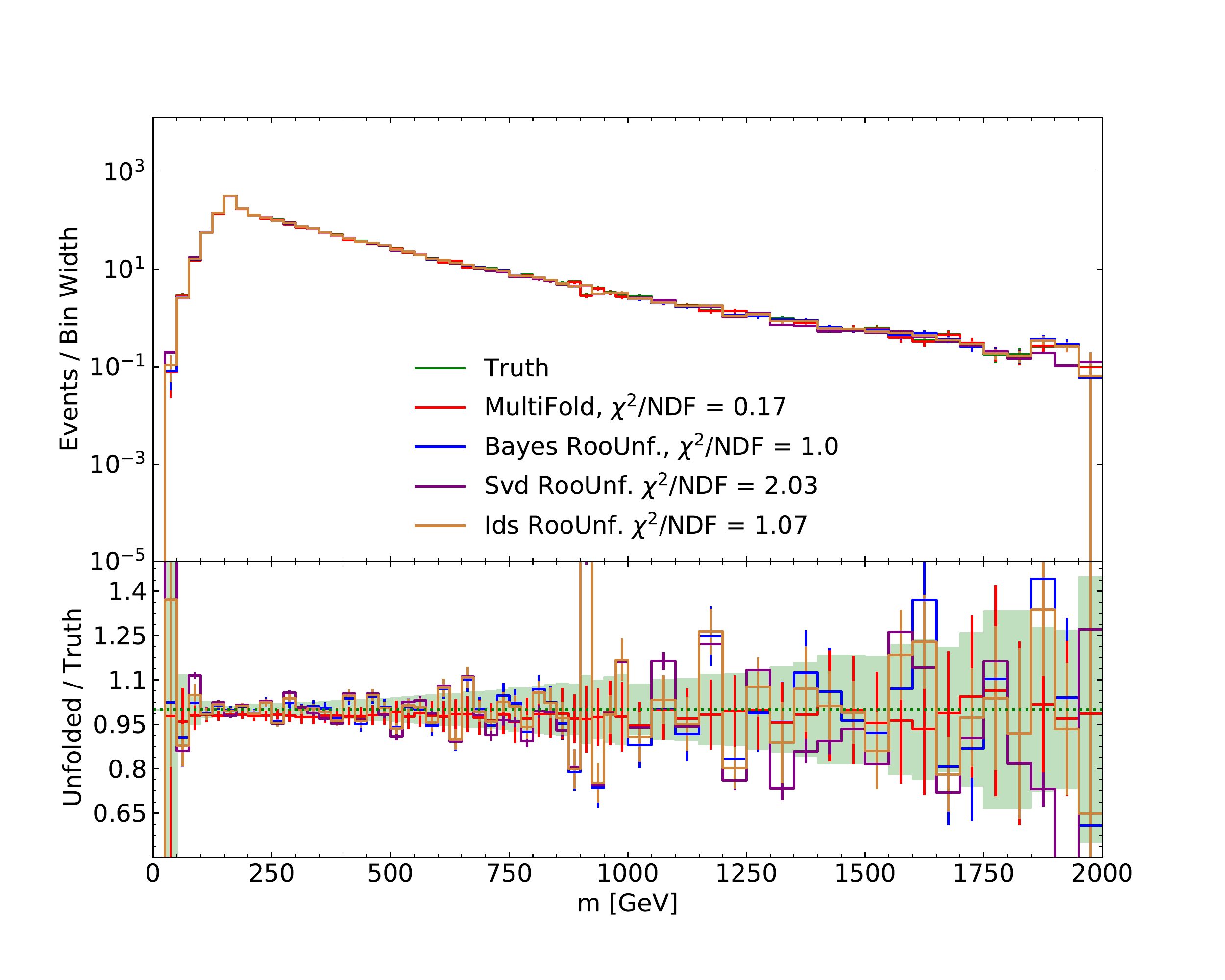}}
  \subfloat[Mass]{\includegraphics[width=0.33\linewidth]{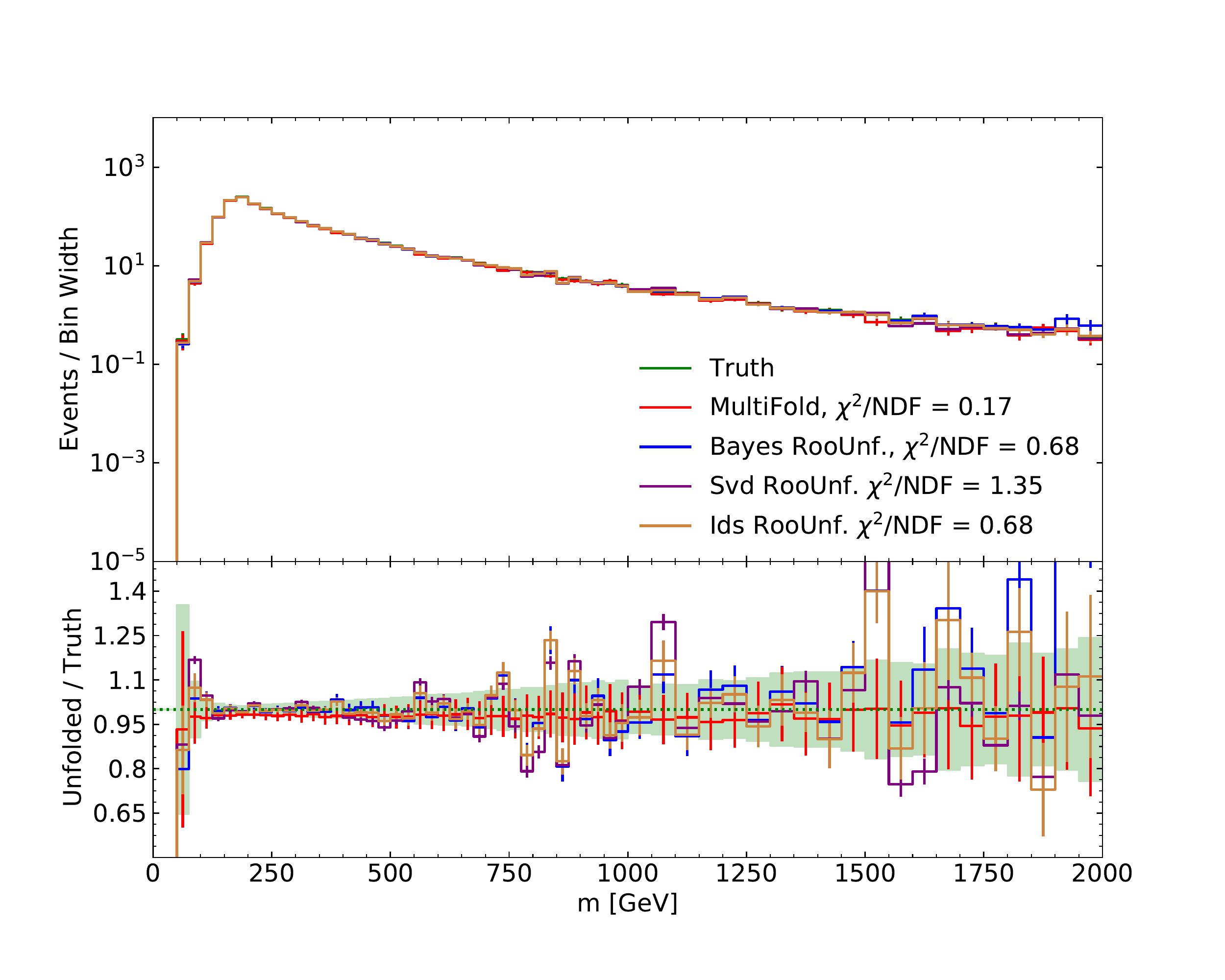}}
  \subfloat[Mass]{\includegraphics[width=0.33\linewidth]{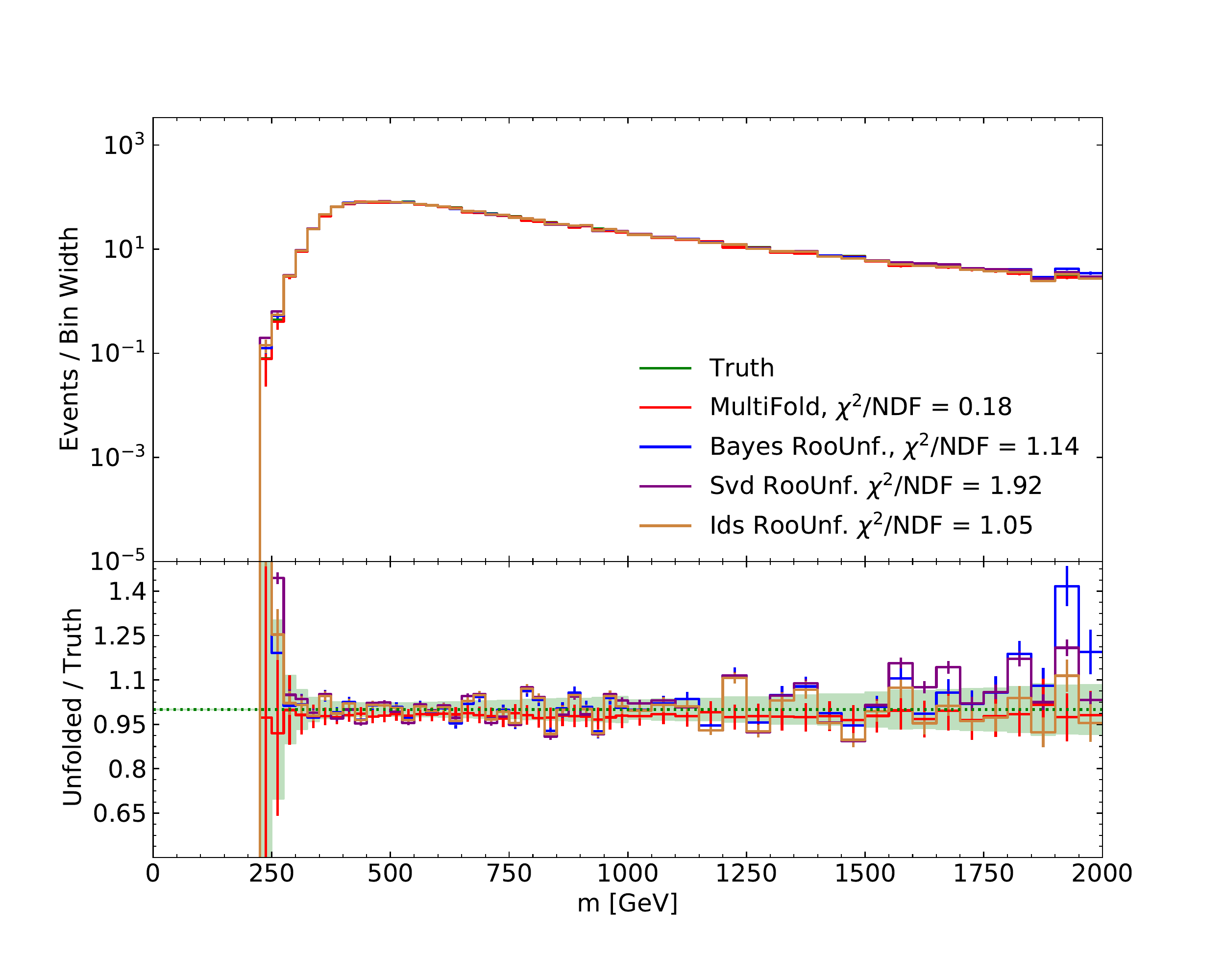}}\\

  \subfloat[Energy]{\includegraphics[width=0.33\linewidth]{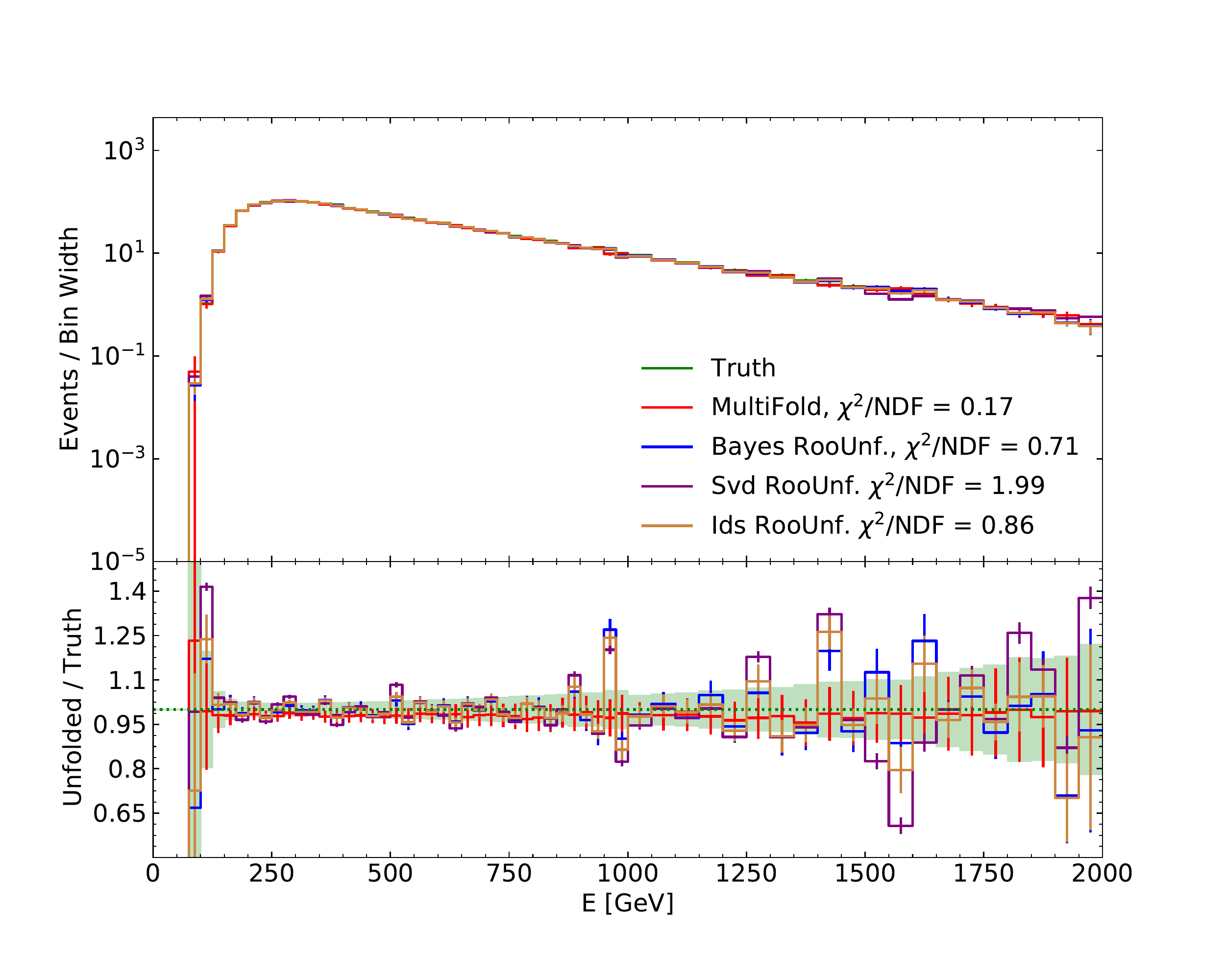}}
  \subfloat[Energy]{\includegraphics[width=0.33\linewidth]{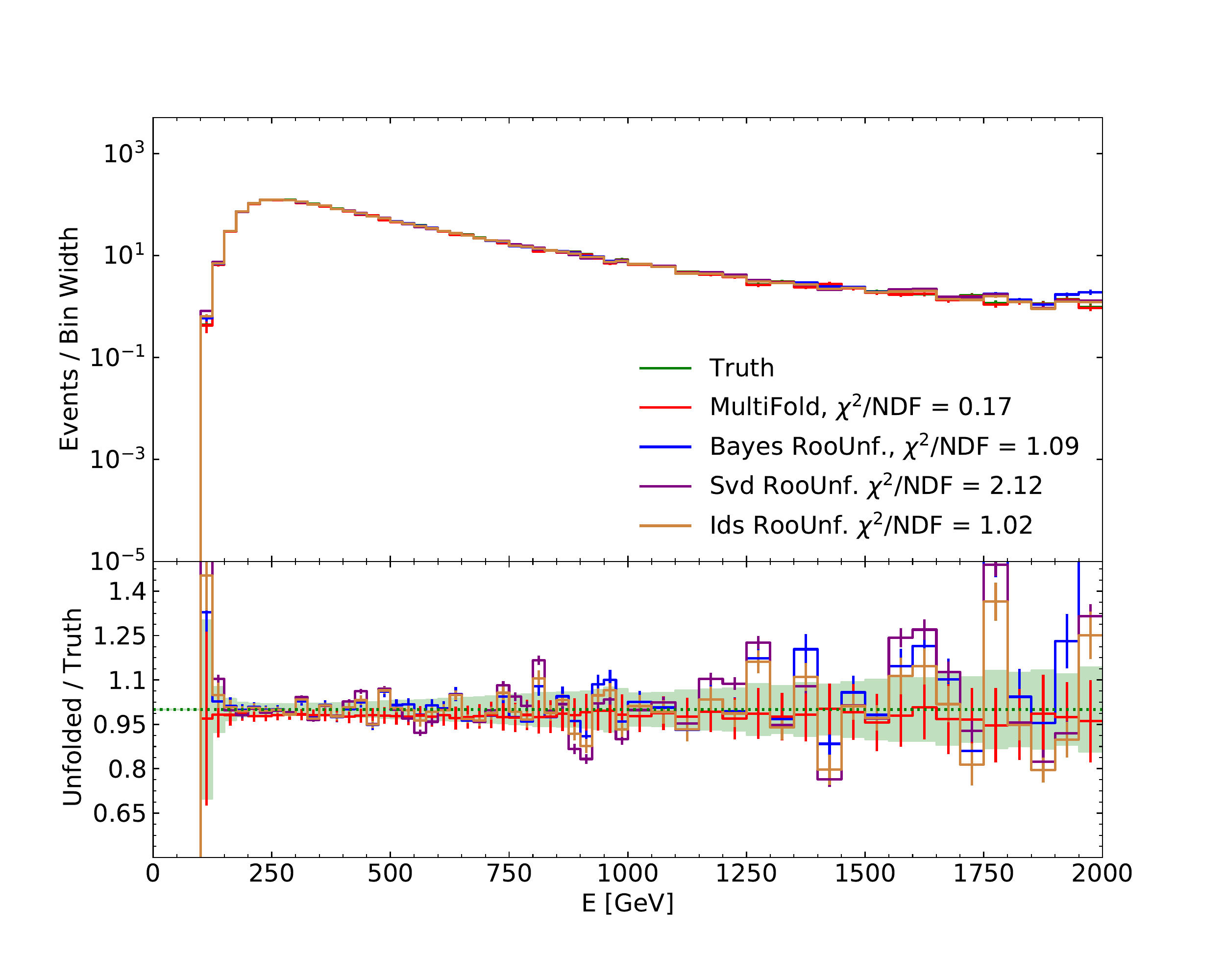}}
  \subfloat[Energy]{\includegraphics[width=0.33\linewidth]{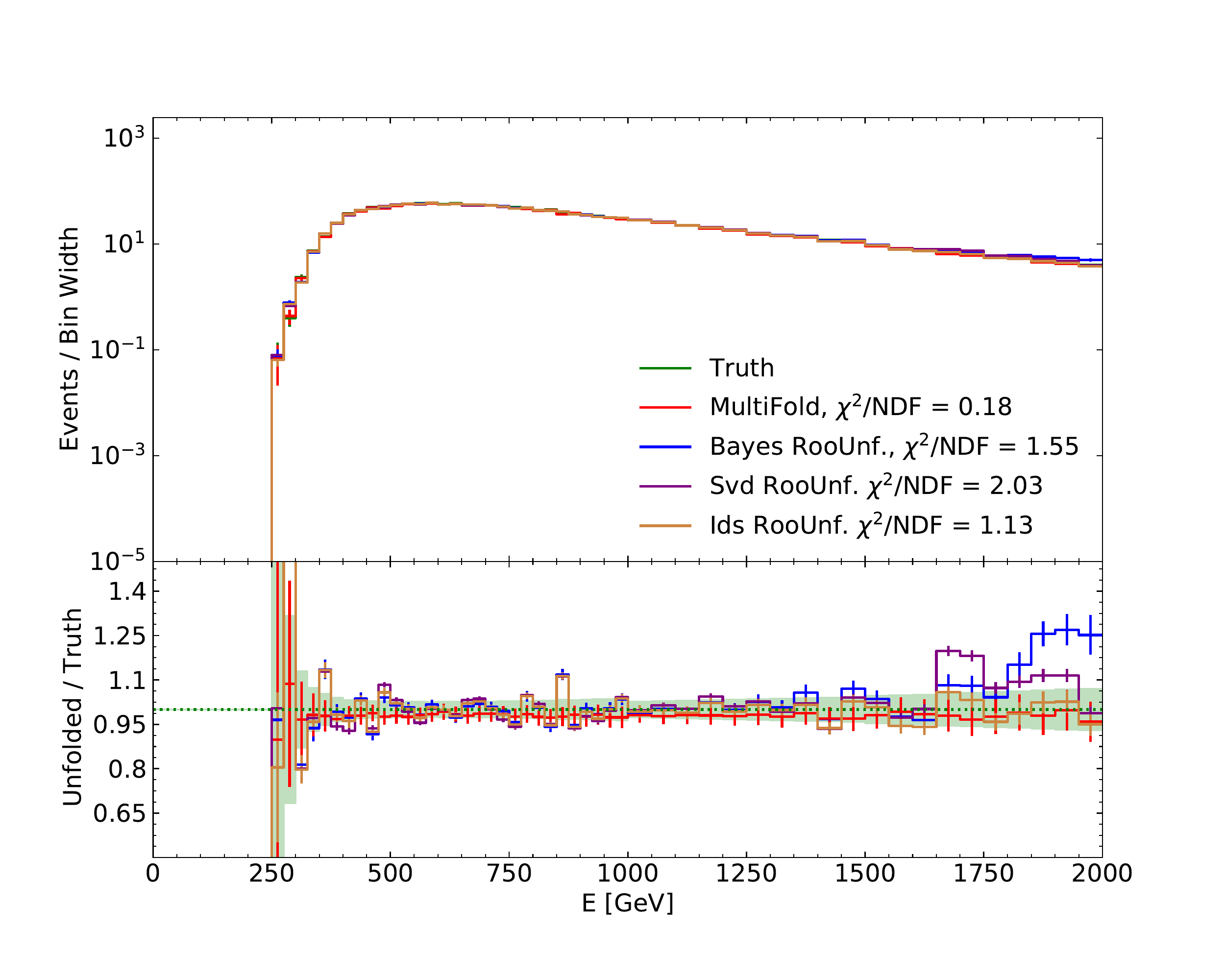}}\\

  \subfloat[Pseudo-rapidity]{\includegraphics[width=0.33\linewidth]{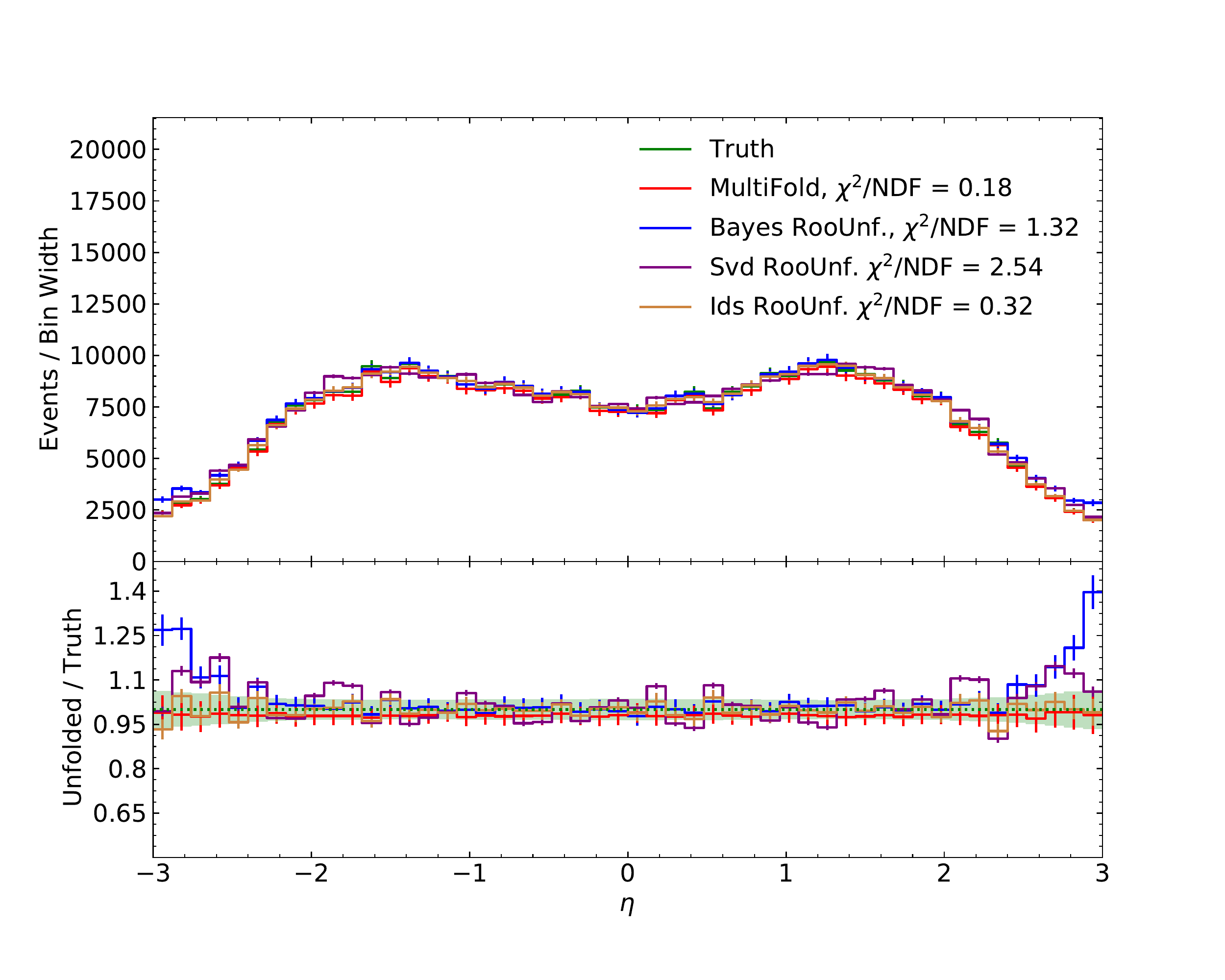}}
  \subfloat[Pseudo-rapidity]{\includegraphics[width=0.33\linewidth]{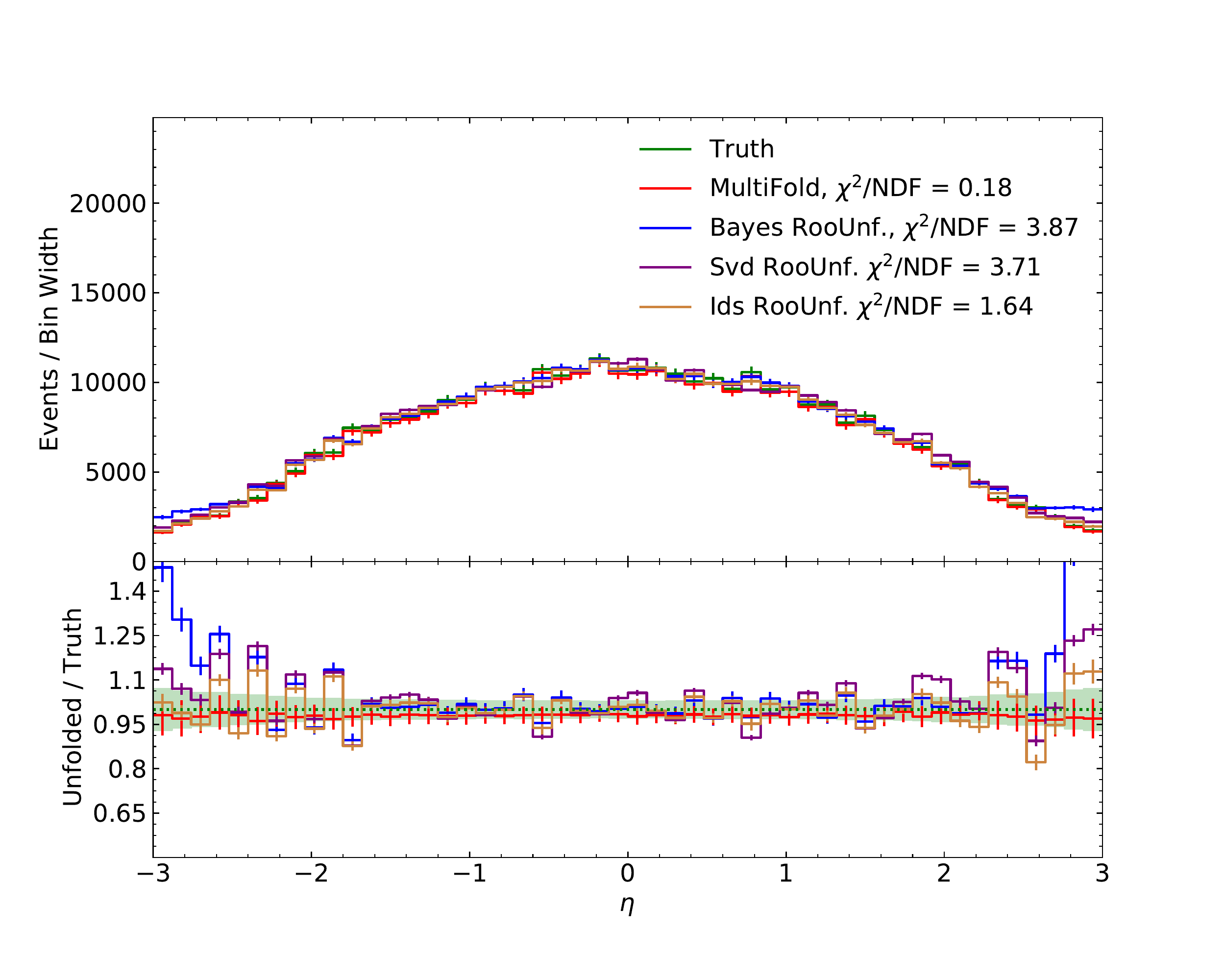}}
  \subfloat[Pseudo-rapidity]{\includegraphics[width=0.33\linewidth]{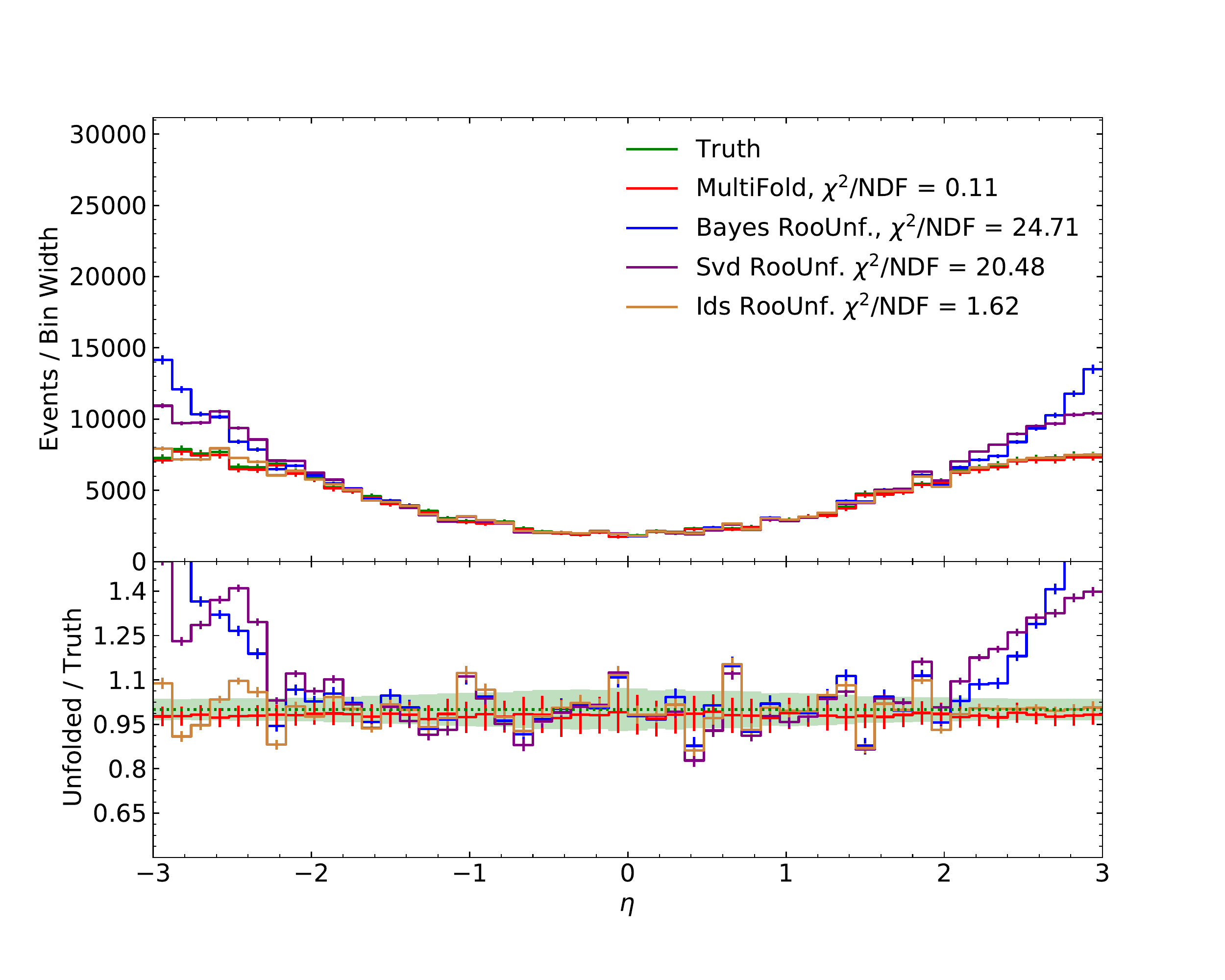}}\\
  \caption{Closure test of the spectra of the hadronically \textbf{- left} leptonically the \textbf{- middle} decaying 
  top quark and of the $t\bar{t}$ system the \textbf{- right} of the $p_T$, mass, energy and pseudo-rapidity 
  with the lower pads showing the ratio of the unfolded to the truth spectrum.}
\end{figure}

\clearpage

\begin{table}[h!]
  \resizebox{\textwidth}{!}{%
  \begin{tabular}{|c|c|c|c|c|}
  \hline
  & Tr. momentum & Mass & Energy & Pseudo-rapidity \\
  \hline
  \hline
  Hadronic top & 0.20 & 0.17 & 0.17 & 0.18 \\
  \hline
  Leptonic top & 0.20 & 0.17 & 0.17 & 0.18 \\
  \hline
  $t\bar{t}$ system & 0.23 & 0.18 & 0.18 & 0.11 \\
  \hline
  \end{tabular}%
  }
  \caption{Results of the closure test, $\chi^2/\mathrm{NDF}$ between the truth and unfolded spectra using the OmniFold method.}
  \label{tab:omni}
  \end{table}

  \begin{table}[h!]
    \resizebox{\textwidth}{!}{%
    \begin{tabular}{|c|c|c|c|c|}
    \hline
    & Tr. momentum & Mass & Energy & Pseudo-rapidity \\
    \hline
    \hline
    Hadronic top & 0.40 & 1.00 & 0.71 & 1.32 \\
    \hline
    Leptonic top & 1.54 & 0.68 & 1.09 & 3.87 \\
    \hline
    $t\bar{t}$ system & 0.56 & 1.14 & 1.55 & 24.71 \\
    \hline
    \end{tabular}%
    }
    \caption{Results of the closure test, $\chi^2/\mathrm{NDF}$ between the truth and unfolded spectra using the Bayes RooUnfold method.}
    \label{tab:bay}
    \end{table}

    \begin{table}[h!]
      \resizebox{\textwidth}{!}{%
      \begin{tabular}{|c|c|c|c|c|}
      \hline
      & Tr. momentum & Mass & Energy & Pseudo-rapidity \\
      \hline
      \hline
      Hadronic top & 1.41 & 2.03 & 1.99 & 2.54 \\
      \hline
      Leptonic top & 2.46 & 1.35 & 2.12 & 3.71 \\
      \hline
      $t\bar{t}$ system & 1.14 & 1.92 & 2.03 & 20.48 \\
      \hline
      \end{tabular}%
      }
      \caption{Results of the closure test, $\chi^2/\mathrm{NDF}$ between the truth and unfolded spectra using the Svd RooUnfold method.}
      \label{tab:svd}
      \end{table}

      \begin{table}[h!]
        \resizebox{\textwidth}{!}{%
        \begin{tabular}{|c|c|c|c|c|}
        \hline
        & Tr. momentum & Mass & Energy & Pseudo-rapidity \\
        \hline
        \hline
        Hadronic top & 0.56 & 1.07 & 0.86 & 0.32 \\
        \hline
        Leptonic top & 1.73 & 0.68 & 1.02 & 1.64 \\
        \hline
        $t\bar{t}$ system & 0.54 & 1.05 & 1.13 & 1.62 \\
        \hline
        \end{tabular}%
        }
        \caption{Results of the closure test, $\chi^2/\mathrm{NDF}$ between the truth and unfolded spectra using the Ids RooUnfold method.}
        \label{tab:ids}
        \end{table}

\section{Conclusion}
Table \ref{tab:omni} summarizes the $\chi^2/\mathrm{NDF}$ results and proves that machine learning 
approach at this particular study performed the best results compared to values in Tables~\ref{tab:bay},~\ref{tab:svd}~and~\ref{tab:ids}.\par
The study aimed to demonstrate the possible potential of the machine learning methods on four typical spectra 
used in high-energy physics. 
A slight disadvantage of the machine learning method might be its initial CPU time needed to train the neural network. 
Although more complex tests of OmniFold unfolding 
need to be performed in the future, the $\chi^2/\mathrm{NDF}$ presents promising results. 

\Acknowledgements
The author gratefully acknowledges the support from the project \\IGA\_PrF\_2021\_004 of Palacky University as well as grant GACR 19-21484S of MSMT, Czech Republic.

\end{document}